\documentclass[12pt]{article}
\usepackage{graphicx}% Include figure files
\usepackage{dcolumn}% Align table columns on decimal point
\usepackage{bm}% bold math
\usepackage{cite}
\usepackage{amsmath}
\newcommand{\be}{\begin{equation}}
\newcommand{\ee}{\end{equation}}
\newcommand{\bea}{\begin{eqnarray}}
\newcommand{\eea}{\end{eqnarray}}
\newcommand{\ba}{\begin{array}}
\newcommand{\ea}{\end{array}}

\catcode`\@=11
\def\lsim{\mathrel{\mathpalette\@versim<}}
\def\gsim{\mathrel{\mathpalette\@versim>}}
\def\@versim#1#2{\vcenter{\offinterlineskip
\ialign{$\m@th#1\hfil##\hfil$\crcr#2\crcr\sim\crcr } }}
\catcode`\@=12

\catcode`@=12
\begin{document}
\thispagestyle{empty}
\begin{flushright}
  IFIC 10-16
\end{flushright}
\vspace{0.3in}

\begin{center}
{\LARGE \bf  Fritzsch neutrino mass matrix from $S_3$ symmetry \\}
\vspace{.5in}
{D.\,Meloni\footnote{Davide.Meloni@physik.uni-wuerzburg.de}, \\
{\small Institut f{\"u}r Theoretische Physik und Astrophysik, \\
Universit{\"a}t W{\"u}rzburg, D-97074 W{\"u}rzburg, Germany}\\
\vspace{.25in}
S.\,Morisi\footnote{morisi@ific.uv.es}, E.\,Peinado\footnote{epeinado@ific.uv.es}}\\
\small AHEP Group, Institut de F\'{\i}sica Corpuscular --
C.S.I.C./Universitat de Val{\`e}ncia \\
\small Edificio Institutos de Paterna, Apt 22085, E--46071 Valencia, Spain 
\end{center}

\begin{abstract}

We present an extension of the Standard Model (SM) based on the discrete flavor 
symmetry $S_3$ which gives a neutrino mass matrix with two-zero texture of Fritzsch-type and
nearly diagonal   charged lepton mass matrix. 
The model is compatible with the normal hierarchy only and predicts 
$\sin^2\theta_{13}\approx 0.01$ at the best fit values of solar and atmospheric parameters 
and maximal leptonic CP violation.

\end{abstract}

\section{\label{int}Introduction}

Although there is a robust evidence that  neutrinos are mixed, many 
aspects of the neutrino physics are not clearly understood yet. Among them, the comprehension of 
the values of the masses and mixing and the differences with respect to the quark sector
are an open problem whose solution seems to be quite far from being found.
Recent data from neutrino oscillations produced the following results:
\bea
0.36 \leq \sin^2 \theta_{23}\leq  0.67 \qquad 0.27 \leq \sin^2 \theta_{12} \leq 0.38 \qquad \sin^2 \theta_{13} <0.053 \,,
\eea
and
\be\ba{c} 2.07\times 10^{-3} ~ eV^2\leq\Delta m^2_{atm}\leq 2.75 \times 10^{-3}~ eV^2,\\
7.03  \times 10^{-5}~ eV^2\leq\Delta m^2_{sol}\leq 8.27 \times 10^{-5}~ eV^2,\ea \ee
at $99.73\%$ confidence level~\cite{data1} (see \cite{data2} for other recent interpretations of the neutrino data).

We have only hints coming from cosmological observations that the absolute values of the neutrino masses should be less than $1~\mbox{eV}$ \cite{Fogli:2008ig}.
In the quark sector the situation is quite different: not only the masses and the hierarchy in the up and down
sectors are better known but also the mixing angles are well measured and strongly differ from the neutrino ones.
A successful ansatz to reproduce these features in the quark sector is the Fritzsch-like 
texture \cite{Fritzsch:1979zq}, where both the up and down quark mass matrices have a simple form
\begin{equation}\label{fr} 
M=\left(
\begin{array}{ccc}
0 &A&0 \\
A^* &C&B \\
0 &B^*&D \\
\end{array}
\right).
\end{equation} 
Such a matrix (already described in, e.g., \cite{others}) gives the well know relation
\begin{equation}\label{t12}
\tan\theta_{12}=\sqrt{\frac{m_1}{m_2}},
\end{equation} 
which predicts the Cabibbo angle whose  small value is a consequence of the strong hierarchy in the masses. 
A texture as in eq.(\ref{fr}) can also be employed for the Majorana neutrino mass matrix;
this is a particular case of the class of two-zero texture \cite{2zeros} which, 
together with the two relations $\Delta m_{atm}=m_3^2-m_1^2$ and $\Delta m_{sol}=m_2^2-m_1^2$, fix the absolute neutrino mass scale as suggested in \cite{Fritzsch:2009vy}. Unlike the quark sector, 
the solar and atmospheric angles can be large due to the fact that in the neutrino sector the hierarchy is not so strong. 

Although a vast class of Fritzsch-like textures (and their phenomenological consequences) 
has been already studied in the literature, in this paper we propose a leptonic model based on the permutation symmetry $S_3$ which naturally gives rise to a Fritzsch-type neutrino Majorana mass matrix 
(and, in addition, to a  nearly diagonal charged leptons). At tree level,  the tau lepton acquires a mass via the spontaneous electroweak symmetry breaking (ESB) driven by one 
$S_3$ doublet and two $S_3$ singlets, whereas the electron and the muon remain massless. Higher order operators, mediated by just one Standard Model (and $S_3$) scalar singlet (called the flavon) are responsible 
for $m_e,m_\mu \ne 0$. In the neutrino sector, the Majorana mass matrix is generated by dimension five  \cite{Weinberg:1979sa} and six operators.

The paper is organized as follows: in the next section we introduce the model; the scalar potential is studied in 
Sec.\ref{scalarp}; the lepton  and neutrino mass matrices are introduced in Secs.\ref{lept} and \ref{neu}, respectively 
whereas their phenomenological consequences are discussed in Sec.\ref{pheno}. Sec.\ref{conc} is devoted to our conclusions.

\section{The model}
\label{themodel}
We propose a model based on $S_3$, the group of permutations of three objects,  which  is the smallest  non-Abelian discrete group. $S_3$ contains one doublet irreducible representation and two singlets. This feature is useful to separate the third family of fermions from the other two and has been  already  used for model building~\cite{s3}. For pioneers papers see \cite{pionerings3} (and
 also references in \cite{otheringroups}).  

The group $S_3$ has two generators $S$ and $T$  satisfying the following
relations:
\be
S^2=T^3=(ST)^2=1~.
\ee
One possible realization is the so-called ``T-diagonal`` basis where  
\be \label{bidimofst}
\ba{lr}
S = \left(\ba{cc} 
0 & 1 \\
1 & 0 
\ea   \right)
&                
T = \left( \ba{cc} 
\omega' & 0 \\
0 & \omega'^2 \\ 
\ea   \right) , 
\ea\ee
with $\omega'=e^{i 2/3 \pi}$.
The tensor products involving pseudo-singlets are given 
by $1'  \times 1' =1 $ and $1' \times 2=2$ while the product of two doublets is  
$2\times 2=2+1+1'$ which, in terms of the components  of the two doublets $A=(a_1,a_2)^{\rm T}$ 
and $B=(b_1,b_2)^{\rm T}$ in the T-diagonal basis, are as follows:
\be \label{tensorprod1s3}
\ba{l}
a_1b_2+a_2b_1 \in 1 \\
a_1b_2-a_2b_1 \in 1' \\
\ea
\qquad
 \left( 
 \ba{c}
 a_2 b_2 \\
 a_1 b_1 \\
  \ea
  \right)
  \in 2\,.
 \ee
The product $2^*\times 2$ is similar to $2 \times 2$ with the exchange of $a_1\leftrightarrow a_2$.

\begin{center}{\it Construction of the model}\end{center}
The Higgs sector is extended from one $SU(2)_L$-doublet to two $SU(2)_L$-doublets, $H_{D}=(H_1,H_2)$ belonging to a doublet 
irreducible representation of $S_3$ and other two $SU(2)_L$ doublets, $H_S$ and $H_S^\prime$, belonging to singlet representations of $S_3$. We also introduce an electroweak scalar singlet $\chi$ 
which turns out to be relevant to give a non-vanishing electron and muon masses. In order to have nearly diagonal charged  lepton mass matrix we assume
two further parity symmetries, so that the global discrete symmetry group of the model is 
$G=S_3 \otimes Z_5 \otimes Z_2$. 
The matter assignment  under $G\otimes SM$ is summarized in Tab.\ref{tab:Multiplet1}.
\begin{table}[h!]
\begin{center}
\begin{tabular}{|c||cccc||ccc|c|}
\hline
fields & $L_D=L_{1,2}$&$L_{3}$ & $l_{R_D}=l_{R_{1,2}}$& $l_{R_{3}}$& $H_{D}$&$H_S$  &$H_S'$& $\chi$ \\
\hline
$SU_L(2)$ & 2& 2& 1&1&2&2&2&1\\
$Y$ & -1& -1& -2& -2 &  1 & 1  & 1  &  0 \\
$S_3$  &  $2$& $1$ & $2$&$1$& $2$&$1$ &$1$ &$1$ \\
$Z_2$ &$+$ &$-$ &$+$&$-$ &$+$ &$+$&$-$&$+$\\
$Z_5$ &$\omega^2$ &$\omega$ &$\omega$&$\omega^2$ &$\omega^3$ &$\omega^4$&$\omega^4$&$\omega^2$ \\
\hline
\end{tabular}
\caption{\it Matter assignment of the model. $\omega$ is the $Z_5$ charge $\omega=e^{i 2/5 \pi}$ and $Y$ is the SM hypercharge in the convention $Y=2(Q-T_3)$, where $T_3$ is the third component of 
the SM $SU(2)$ doublets. }
\label{tab:Multiplet1}
\end{center}
\end{table}

\section{The scalar potential}
\label{scalarp}
The most general Higgs potential invariant under $G\times SM$ is as follows:%\footnote{other terms invariant under $G$ are omited because are reabsorbed in other, for instance $(H_D^\dagger H_D^\dagger)_1(H_DH_D)_1\sim(H_D^\dagger H_D)_D(H_D^\dagger H_D)_D$}:
\be
\ba{lcl}
V&=&\mu_1 H^{\prime \dagger}_SH^\prime_S +\mu_2 (H_D^\dagger H_D)_1+\mu_3H^\dagger_SH_S+\mu_4 |\chi|^2+\lambda_1 |\chi|^4\\\\ && +
(\lambda_2H_D^\dagger H_D +\lambda_3 H^{ \dagger}_SH_S+ \lambda_4H^{\prime \dagger}_SH^\prime_S  )|\chi|^2+\lambda_5 [(H_D^\dagger H_D)]^2+\lambda_6 [(H_D^\dagger H_D)_{1^\prime}]^2\\\\ && +
\lambda_7[(H_D^\dagger H_D)_2]^2+\lambda_7^\prime(H_D^\dagger H_D^\dagger)_1(H_DH_D)_1 +\lambda_8(H^\dagger_SH_S)^2\\\\ 
&&+\lambda_9'(H_D^\dagger H_D)_1H^{\prime \dagger}_SH^\prime_S
+\lambda_9''(H_D^\dagger H'_S)_2(H^{\prime \dagger}_SH_D)_2
+\lambda_9'''((H_D^\dagger H'_S)_2^2+h.c.)
+\\\\ &&
+\lambda_{10}'(H_D^\dagger H_D)_1H^\dagger_SH_S+ 
\lambda_{10}''(H_D^\dagger H^\prime_S)_2(H^{\prime \dagger}_SH_D)_2
+\lambda_{10}'''((H_D^\dagger H_S)_2^2+h.c.)+
\\\\ &&+
\lambda_{11}(H_D^\dagger H_D^\dagger)_2(H_DH_D)_2+\lambda_{12} (H^{\prime \dagger}_SH^\prime_S)^2+\\\\ &&+{\lambda_{13}}' H^{\prime \dagger}_SH^\prime_SH^{ \dagger}_SH_S+{\lambda_{13}}'' (H^{\prime \dagger}_SH^{\prime \dagger}_SH_SH_S+h.c.)+
{\lambda_{13}}''' H^{\prime \dagger}_SH_SH^{ \dagger}_SH'_S
\ea
\ee
where we used the subscripts $1,1^\prime$ and $2$ to refer to the $S_3$ contractions  when necessary  and, for any Higgs fields, $\tilde H=-i \tau_2^T H^*$.
In the case of real vev's, that is
\be
\ba{l}
\langle H_D \rangle=(v_1,~v_2),\\
\langle H_S \rangle=v_S,\\
\langle H^\prime_S \rangle=v^\prime_S,\\
\langle \chi \rangle=v_\chi,
\ea
\ee
the potential can be written as\footnote{Where $\lambda_i=\lambda_i'+\lambda_i''+\lambda_i'''$ for $i=9,10,13$.}
\begin{equation}\begin{array}{lcl}
V&=&(\lambda_{11}+\lambda_5+\lambda_6)(v_1^4 +v_2^4)+v_S^4 \lambda_8+v_S^{\prime 4} \lambda_{12}+v_S^2 v_S^{\prime 2} \lambda_{13}+v_S^{\prime 2} \mu_1+v_S^2 \mu_3+\\&+&\left( \mu_4+v_S^2 \lambda_3+v_S^{\prime 2} \lambda_4 \right) \chi^2+
\lambda_1 \chi ^4+\left(v_S^2 \lambda_{10}+v_S^{\prime 2} \lambda_9+\mu_2+\lambda_2 \chi ^2\right)(v_2^2+v_1^2) +\\&+& \left(2 v_2^2v_1^2 (\lambda_5-\lambda_6+\lambda_7)\right).
\end{array}
\end{equation}
The minima of $V$ are found solving the minimizing equations:
\be
\ba{l}
\frac{\partial V}{\partial v_1}=2 v_1 \left[\lambda_2 v_\chi^2+\lambda_{10}v_S^2+\lambda_9 v^{\prime 2}_S+2 v_1^2 (\lambda_{11}+\lambda_5+\lambda_6)+2 v_2^2 (\lambda_5-\lambda_6+\lambda_7)+\mu_2\right]=0,\\
\\
\frac{\partial V}{\partial v_2}=2 v_2 \left[\lambda_2 v_\chi^2+\lambda_{10}v_S^2+\lambda_9 v^{\prime 2}_S+2 v_2^2 (\lambda_{11}+\lambda_5+\lambda_6)+2 v_1^2 (\lambda_5-\lambda_6+\lambda_7)+\mu_2\right]=0,\\
\\
\frac{\partial V}{\partial v_S}=2v_S \left[\lambda_3  v_\chi^2+\lambda_{10} \left(v_1^2+v_2^2\right)+2v_S^2 \lambda_8+v^{\prime 2}_S \lambda_{13}+\mu_3\right]=0,\\
\\
\frac{\partial V}{\partial v^\prime_s}=2 v^\prime_s \left[\lambda_4 v_\chi^2+\lambda_9 \left(v_1^2+v_2^2\right)+2 v^{\prime 2}_S \lambda_{12}+v_S^2 \lambda_{13}+\mu_1\right]=0,\\
\\
\frac{\partial V}{\partial v_\chi}=
2 v_\chi \left[\mu_4+2 \lambda_1 v_\chi^2+\lambda_2 \left(v_1^2+v_2^2\right)+\lambda_3v_S^2+\lambda_4 v^{\prime 2}_S\right]=0 \,.
\ea
\ee
The second equation is satisfied for $v_2=0$. From the remaining equations we can easily 
get the vevs of the other scalars in terms of the couplings of the Higgs potential; in particular, a solution with $v_1\ne 0 $ can be found 
and the  vev alignment of the  $S_3$ Higgs doublet assumes the structure:
\begin{equation}\label{vev}
\langle H_D \rangle=(v,0) \,.
\end{equation}
For this vev configuration, it is possible to find a huge region of the Higgs parameter space where the eigenvalues of the Hessian of the potential are all positive and therefore where 
the Higgs potential has a local minimum. Note that a solution of the form $\langle H_D \rangle=(0,v)$ is physically equivalent to eq.(\ref{vev}), producing the same phenomenology in the charged lepton and neutrino 
sectors. In fact it corresponds to the exchange of $L_1$ with $L_2$. 
We also verified numerically that, in the large parameter space where  eq.(\ref{vev}) is a minimum, other solutions like $\langle H_D \rangle=v\,(1,1)$ do not produce positive definite Hessian.
The mass spectra of the Higgs particles will be discussed elsewhere. 
\section{Leptons}
\label{lept}
The most general Lagrangian invariant under $G\times SM$ is given by:
\begin{eqnarray}
\mathcal{L}_{}&=&
\frac{y_1}{\Lambda} \overline{L_D}H_Dl_{R_D}\chi^*+\frac{y_2}{\Lambda} \overline{L_D}H_Sl_{R_D}\chi
+y_3  \overline{L_3}H_Sl_{R_3}\, ,
\end{eqnarray}
where $\Lambda$ is the cut-off scale.
Higher order terms only appear at ${\cal O}(1/\Lambda^2)$  and will be considered negligible for our discussion.
% The charge assignment of the flavon field $\chi$ under $Z_2$ and $Z_5$ prevents higher 
% order terms to modify the coupling $y_1, y_2$ and $y_3$ at the next to leading 
% order, the first useful corrections being at ${\cal O}(1/\Lambda^2)$. At the same order, $1/\Lambda^2$, we expect new terms (not modifying the previous
%  Yukawas) to appear in the Lagrangian, that will be considered negligible for 
% our discussion. 
From eq.(\ref{vev}) the charged lepton mass matrix is: 
\begin{equation} 
M_l=\left(
\begin{array}{ccc}
\frac{y_2}{\Lambda} v_S v_\chi& 0&0\\
\frac{y_1}{\Lambda} v v_\chi&\frac{y_2}{\Lambda} v_S v_\chi&0\\
0&0&y_3 v_S\\
\end{array}
\right).
\label{chargedlept}
\end{equation} 
When $v_\chi$ is equal to zero only the $\tau$ lepton is massive. The electron and muon masses are generated
by the vev of the scalar $\chi$ and are then suppressed by the large scale $\Lambda$. %By introducing
%a Froggatt Nielsen symmetry $U_{FN}(1)$ such that only the right-handed
%fields are charged, we expect 
%\begin{equation}\label{scale}
%y_1\sim y_2\sim y_3\ll y_4.
%\end{equation}
%The condition in eq.(\ref{scale}) implies that $M_l$ is hierarchical and
%the diagonalizing matrix is very close to the identity. 
The matrix $M_l M^\dagger_l$ has three distinct eigenvalues that can be identified with the squared charged fermion masses as:
\begin{eqnarray}
 m_e^2&=&\frac{\varepsilon^2}{2}\, \left(v^2 y_1^2 + 2 v_S^2 y_2^2 - 
   v y_1 \sqrt{v^2 y_1^2 + 4 v_S^2 y_2^2}\right) \nonumber \\
m_\mu^2&=&\frac{\varepsilon^2}{2}\, \left(v^2 y_1^2 + 2 v_S^2 y_2^2 +
   v y_1 \sqrt{v^2 y_1^2 + 4 v_S^2 y_2^2}\right)  \\
m_\tau^2 &=& v_S^2 y_3^2 \nonumber
\end{eqnarray}
where we introduced the short-hand notation $\varepsilon = v_\chi/\Lambda$. We see that for $\varepsilon \ll 1$, the hierarchy among the $\tau$ and the lightest charged leptons
is easily reproduced although the latter, in absence of any fine-tuning among the Yukawas and/or the Higgs vevs, are expected to be of the same order of magnitude. 
We address this question in the next section.
The mass matrix for the charged leptons can be written in terms of the physical lepton masses as:
\be
\label{emmeelle}
M_l=
\left(
\ba{ccc}
\sqrt{m_e m_\mu}& 0 & 0 \\
-m_\mu (1-\frac{m_e}{m_\mu}) & \sqrt{m_e m_\mu} & 0 \\
0 & 0 & m_\tau
\ea
\right),
\ee
and the squared matrix $M_l M^\dagger_l$ is then diagonalized by:
\be
U_L=\left(\ba{ccc}
\frac{1}{\sqrt{1+\frac{m_e}{m_\mu}}} & -\sqrt{\frac{m_e}{m_\mu}}\frac{1}{\sqrt{1+\frac{m_e}{m_\mu}}} &0 \\
\sqrt{\frac{m_e}{m_\mu}}\frac{1}{\sqrt{1+\frac{m_e}{m_\mu}}} & \frac{1}{\sqrt{1+\frac{m_e}{m_\mu}}} & 0 \\
0 & 0 & 1
\ea\right) \sim 
\left(
\ba{ccc}
1 & -0.07 &0 \\
0.07 & 1 & 0 \\
0 & 0 & 1
\ea\right)\,.\label{ul}
\ee

\section{Neutrino}
\label{neu}
The neutrino masses are generated by non-renormalizable operators of dimension 5 and 6  invariant under the group $G\times SM$\footnote{Dimension 7 operators can be built, for instance, adding 
the singlet $\chi^2$ or doublets $H^\dagger H$ to the previous $d=5$ operators and will then be neglected.}:
\begin{eqnarray}
\Lambda \cdot \mathcal{L}_{\nu}&=&
y^\nu_1(\overline{L}_D\overline{L}_D)_{1}(\tilde{H_D}\tilde{H_D})_{1}+
y^\nu_2(\overline{L}_D\overline{L}_D)_2(\tilde{H_D}\tilde{H_D})_2+
y^{\nu}_3(\overline{L}_D\tilde{H_D})_1(\overline{L}_D\tilde{H_D})_{1}+ \nonumber  \\
\nonumber
&&y^{\nu}_4(\overline{L}_D\tilde{H_D})_{1'}(\overline{L}_D\tilde{H_D})_{1'}+
y^{\nu}_5(\overline{L}_D\tilde{H_D})_{2}(\overline{L}_D\tilde{H_D})_{2} + \nonumber\\&&
{y^\nu_6}'(\overline{L}_D\overline{L}_D)_{1}(\tilde{H_S}\tilde{H_S})_1\,\chi^*/\Lambda+
{y^\nu_6}''(\overline{L}_D\overline{L}_D)_{1}(\tilde{H_S'}\tilde{H_S'})_1\,\chi^*/\Lambda +  \\ &&
\nonumber  
y^\nu_7\overline{L}_3\overline{L}_3(\tilde{H_D}\tilde{H_D})_1\,\chi/\Lambda+
{y^\nu_8}'\,\overline{L}_3\overline{L}_3 \tilde{H_S}\tilde{H_S}+{y^\nu_8}''\,\overline{L}_3\overline{L}_3 
\tilde{H_S'}\tilde{H_S'}+ \nonumber\\&& 
y^\nu_{9}(\overline{L}_D\tilde{H_D})_1 \overline{L}_3\tilde{H_S'} \nonumber \,, \label{lagneu}  
\end{eqnarray}
where we assumed that  the large energy scale which suppresses these operators is of the same order of the cutoff scale $\Lambda$.
The only operators of dimension six are those proportional to $y^\nu_6={y^\nu_6}'+{y^\nu_6}''$ and 
$y^\nu_7$.  
%Notice also that we grouped into the $y^\nu_6$ and $y^\nu_8$ terms those operators which will just give a redefinition of the Yukawas.
From eq.(\ref{lagneu}) the neutrino mass matrix is as follows:

\begin{equation}\label{Mnu} 
M_\nu=\left(
\begin{array}{ccc}
0 & 2\,y^\nu_6 (v_S^2+v_S'^2)v_\chi/\Lambda &0 \\
 2\,y^\nu_6 (v_S^2+v_S'^2)v_\chi/\Lambda&(y^\nu_2+y^\nu_3+y^\nu_4) v^2 & y^\nu_9 v v_S' \\
0&y^\nu_9 v v_S'& y^\nu_8 (v_S^2+v_S'^2)\\
\end{array}
\right)\equiv
\left(
\begin{array}{ccc}
0 &b&0 \\
 b&a&c\\
0&c&d\\
\end{array}
\right)\,,
\end{equation}
where $y^\nu_8={y^\nu_8}'+{y^\nu_8}''$.
Before discussing the phenomenological consequences of such a matrix, it is useful to  get an estimate of the relevant Yukawa parameters and a relation among the vevs $v$ and $v_S$. 
Comparing eqs.(\ref{chargedlept}) and (\ref{emmeelle}) and using the parameterization in eq.(\ref{Mnu}) we get:
\begin{eqnarray}
\sqrt{m_e m_\mu}&=&y_2\,v_S \,\varepsilon \nonumber \\
m_\mu \left(1-\frac{m_e}{m_\mu}\right) &=& y_1\,v \,\varepsilon  \nonumber \\
m_\tau&=&y_3\,v_S  \nonumber   \\
a &=& y^\nu \,v^2  \nonumber \\
b &=& 2\,y_6^\nu (v_S^2+ v_S^{'2}) \, \varepsilon \nonumber \,,
\end{eqnarray}
where we assumed that $y_2^\nu+y_3^\nu+y_4^\nu=y^\nu$.
We assume 
$$
v\gsim v_s \sim \mathcal{O}(100)\,\text{GeV},\qquad \varepsilon\sim \mathcal{O}(10^{-2})
$$
then if 
$$
y_1\sim \mathcal{O}(10^{-1}),\qquad y_2\sim \mathcal{O}(10^{-2}),\qquad y_2\sim \mathcal{O}(10^{-2}),\qquad
$$
we have the correct charged lepton mass hyerarchies. 
We then consider the $b/a$ ratio 
%and extract $\varepsilon$ from the first of the previous equations, obtaining:
%\begin{eqnarray}
% \frac{b}{a}&=& \frac{2\,y_6^\nu \sqrt{m_e m_\mu}(v_S^2+ v_S^{'2})}{y^\nu\, y_2 \,v^2\,v_S}
%\end{eqnarray}
\begin{eqnarray}
 \frac{b}{a}&=& \frac{2\,y_6^\nu \varepsilon (v_S^2+ v_S^{'2})}{y^\nu \,v^2\,v_S}.
\end{eqnarray}
%
%
%The vev $v$ can be obtained from the second equation and, further assuming  $v_S^{'} \ll v_S$ 
%we arrive at the final expression\footnote{Since there are more observables than parameters, 
%several assumptions can be made among the various vevs.}:
%\begin{eqnarray}
% \frac{y_1^2\,y_6^\nu}{y^\nu\,y_2^3}\,\left(\frac{a}{b}\right) \sim 10^4 \, (v_S/GeV)\,.
%\end{eqnarray}
%%
%
We numerically verified that $\left(\frac{a}{b}\right)\sim {\cal O}(1)$ so %that, assuming $v_S \sim {\cal O} (100)$ GeV,
the Yukawa parameters must satisfy
$$
y_6^\nu\sim\mathcal{O}(1),\quad y^\nu\sim\mathcal{O}(10^{-2}).
$$
%
%
%\begin{eqnarray}
% \frac{y_1^2\,y_6^\nu}{y^\nu\,y_2^3}\  \sim 10^6  \,.
%\end{eqnarray}
%The equality can be easily accomodated for several choices of the Yukawas. In particular, one can choose
%\begin{eqnarray}
% y_2  \sim {\cal O} (0.1)  \qquad y_1 \sim {\cal O}(10)
%\end{eqnarray}
%from which
%\begin{eqnarray}
% v  \sim  10 \, \, GeV  \qquad \varepsilon \sim 7 \cdot 10^{-4}\qquad \frac{y_6^\nu}{y^\nu} \sim {\cal O} (10)\,.
%\end{eqnarray}
%
With these assumptions, the hierarchy in the charged leptons is recovered, higher order terms with more that one flavon insertions can be safely neglected and 
the largest vev is generated by the $S_3$ singlet Higgs $H_S$ that can be identified with the Standard Model Higgs.

The mass matrix in eq.(\ref{Mnu}) depends on five real parameters, one of which is related to 
the Dirac phase.
The other four parameters  can be fixed using the experimental information from both 
solar and atmospheric sectors, namely the solar and atmospheric 
mixing angles and squared mass differences. 
The model allows for correlations among the angle $\theta_{13}$ and the CP phase 
$\delta$ that can be easily obtained using  
the zeros of the Fritzsch texture.
%\footnote{Notice that the filling of the (11) element can only be achieved 
%adding  $d=7$ operators to the lagrangian. However, since $vev/\Lambda \sim {\cal O}(10^{-1})$, the relative correction
%is smaller than the charged lepton contribution and can be safely neglected.} 
The previous mass matrix is
diagonalized by a unitary matrix $U^\nu$ as 
\be
U^{\nu T} M_\nu U^\nu =\mbox{diag}(\mu_1,\mu_2,\mu_3)
\ee
where $\mu_i=m_i e^{i\phi_i}$ and $\phi_i$ are Majorana phases.
Writing $U^\nu$ in the CKM-like form\footnote{We have used the short-hand notation $s^\nu_{ij}=\sin \theta_{ij}^\nu$ and  $c^\nu_{ij}=\cos \theta_{ij}^\nu$.} 
\be\label{Usp}
U^\nu=\left(
\begin{array}{ccc}
 c^\nu_{12}c^\nu_{13}& c^\nu_{13}s^\nu_{12}& e^{-i \delta_\nu} s^\nu_{13}\\
 -c^\nu_{23}s^\nu_{12}-c^\nu_{12}e^{i \delta_\nu} s^\nu_{13}s^\nu_{23}& c^\nu_{12}c^\nu_{23}-e^{i \delta_\nu} s^\nu_{12}s^\nu_{13}s^\nu_{23}& c^\nu_{13}s^\nu_{23}\\
 -c^\nu_{12}c^\nu_{23}e^{i \delta_\nu} s^\nu_{13}+s^\nu_{12}s^\nu_{23}& -c^\nu_{23}e^{i \delta_\nu} s^\nu_{12}s^\nu_{13}-c^\nu_{12}s^\nu_{23}& c^\nu_{13}c^\nu_{23}
\end{array}
\right),
\ee
and using the fact that the elements $(M_\nu)_{11}$ and $(M_\nu)_{13}$ are zero (see eq.\,(\ref{Mnu})), 
eq.(\ref{Usp}) implies:

\be
\ba{l}
\mu_2= \mu_1\frac{\cos \theta^\nu_{12} 
\left(-\cot \theta^\nu_{12} \cos \theta^\nu_{23}+\sin \theta^\nu_{12} \sin \theta^\nu_{13} \sin \theta^\nu_{23}e^{i \delta_\nu } \right)}
{\cos \theta^\nu_{23} \sin \theta^\nu_{12}+\cos \theta^\nu_{12}  \sin \theta^\nu_{13} \sin \theta^\nu_{23}e^{i \delta_\nu }},\\
\\
\mu_3= - \mu_1\frac{\cos \theta^\nu_{12} \cos^2 \theta^{\nu}_{13}  \sin \theta^\nu_{23}e^{-i \delta_\nu }}
{\sin \theta^\nu_{13} \left(\cos \theta^\nu_{23} \sin \theta^\nu_{12} +\cos \theta^\nu_{12} \sin \theta^\nu_{13} \sin \theta^\nu_{23}e^{i \delta_\nu } \right)}\,.
\ea
\label{massrel1}
\ee
Our model is compatible with the normal mass ordering only because the ratio $|\mu_2|^2/|\mu_3|^2$ is always less than $1$; expanding it up to second order in $\sin \theta_{13}^{\nu}$ we get:
\be
\frac{|\mu_2|^2}{|\mu_3|^2} = \cot^2 \theta^{\nu}_{12}\, \cot^2 \theta^{\nu}_{23}\,  \sin^2\theta^{\nu}_{13}+ {\cal O} 
(s^{\nu 3}_{13})\,,
\ee
and we checked that higher order corrections do not modify our statement. 
The mass differences are written as:
\be
\Delta m^2_{sol}=m_1^2 \;\left[\frac{\cos \theta^{\nu}_{23} (\cos \theta^{\nu}_{12} \cos \theta^{\nu}_{23} \csc^2 \theta^{\nu}_{12}-2 \cos \delta^{\nu} \cot \theta^{\nu}_{12} \sin \theta^{\nu}_{13} \sin \theta^{\nu}_{23})}{D_\nu}\right],
\label{dsol}\ee
and
\be
\Delta m^2_{atm}=m_1^2 \left(-1+\frac{\cos^2 \theta^{\nu}_{12} \cos^2 \theta^{\nu}_{13}\cot^2 \theta^{\nu}_{13} \sin^2 \theta^{\nu}_{23}}{D_\nu}\right),
\label{datm}
\ee
where 
\bea
D_\nu&=&\cos^2 \theta^{\nu}_{23} \sin^2 \theta^{\nu}_{12}+  \sin 2\theta^{\nu}_{12}  \sin 2\theta^{\nu}_{23}\sin \theta^{\nu}_{13}\cos \delta^{\nu}/2 +\cos^2 \theta^{\nu}_{12} \sin^2 \theta^{\nu}_{13} \sin^2 \theta^{\nu}_{23}\,.
\eea
From the  ratio $\alpha=\Delta m^2_{sol}/\Delta m^2_{atm}$ we find a relation between $\alpha$ and the mixing angles $s^{\nu }_{12}$, $s^{\nu }_{13}$ and $s^{\nu}_{23}$
\be
\label{alpha}
\alpha = 2 \sin^2 \theta_{13}^\nu \cot \theta^\nu_{23} 
\csc \theta^\nu_{12} \sec \theta^\nu_{12} (-\sin \theta_{13}^{\nu} \cos \delta_\nu + \cot 2 \theta^\nu_{12} \cot \theta^\nu_{23} )\,.
\ee
which will be used below to constrain the physical $\theta_{13}$ and $\delta$.

\section{Phenomenology} 
\label{pheno}
To study the phenomenological implication of our model, it is necessary to relate the parameters in eq.(\ref{Usp})
to the physical ones. This can be achieved introducing the rotations from the charged lepton sector
described in Sec.\ref{lept}; the resulting mapping is a set of implicit relations that are quite cumbersome and will not be explicitly presented here. We limit ourselves to describe the procedure which allows us to extract the predictions of our model.
The lepton mixing  is defined by $V=U_L^{\dagger}U^\nu$ and  we can write %$U^\nu$ as 
\be\label{Unu}
U^\nu=U_L  V\,,
\ee
where $V$ is parametrized in the standard form as in   eq.\,(\ref{Usp}) replacing $s_{ij}^\nu$ and $c_{ij}^\nu$ with
the physical (that is measurable) $s_{ij}=\sin\theta_{ij}$ and $c_{ij}=\cos\theta_{ij}$, respectively. 
Taking the ratio of $U^\nu_{23}$ and $U^\nu_{33}$ from eq.\,(\ref{Unu}) 
we find an expression for $s^\nu_{23}$ in terms of the physical parameters $\theta_{12}$, $\theta_{23}$, $\theta_{13}$ 
and the  Dirac phase $\delta$ (and the corrections from the charged leptons). 
In the same way, always using eq.\,(\ref{Unu}),  we can express $s^\nu_{12}$ and $s^\nu_{13}$ as a function of  
$\theta_{12}$, $\theta_{23}$, $\theta_{13}$, $\delta$; finally, $\delta_\nu$ is the argument of the element $(13)$
of the matrix $U_L V$. In this way we have all the parameters $\theta_{12}^\nu$, $\theta_{13}^\nu$, $\theta_{23}^\nu$ and $\delta_\nu$ as a function of the neutrino mixing angles $\theta_{13}$, $\theta_{12}$, $\theta_{23}$ and the  phase $\delta$. These relations can be inserted into eq.\,(\ref{alpha}) to get an implicit connection among 
the mixing parameters and $\alpha$, which is a characteristic of our model.
Also the lightest mass eigenstate can be related to the same parameters and  $\Delta m^2_{atm}$
using eq. (\ref{datm}). 

\begin{center} 
\begin{figure}[h!] 
\includegraphics[width=8cm]{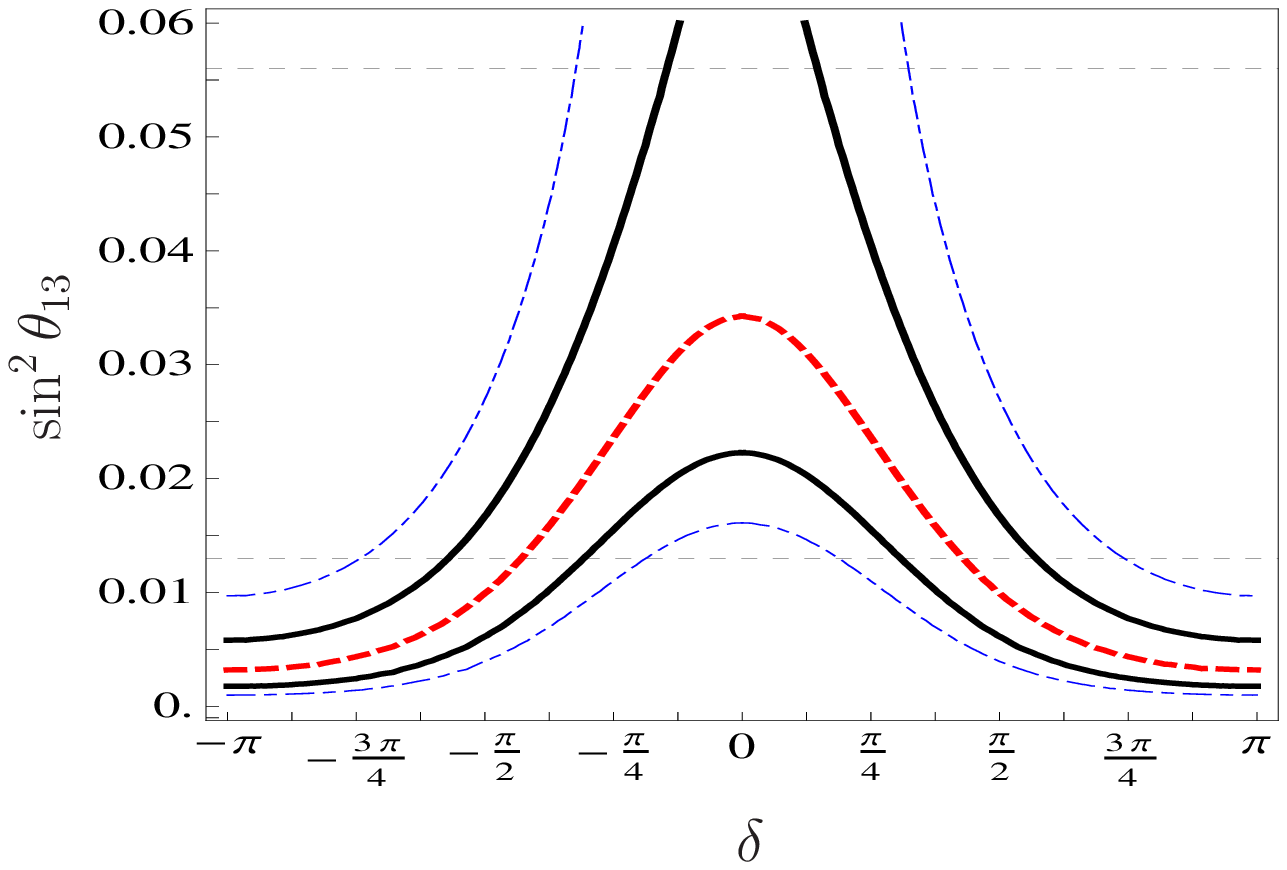}\hspace{.5cm}\includegraphics[width=8cm]{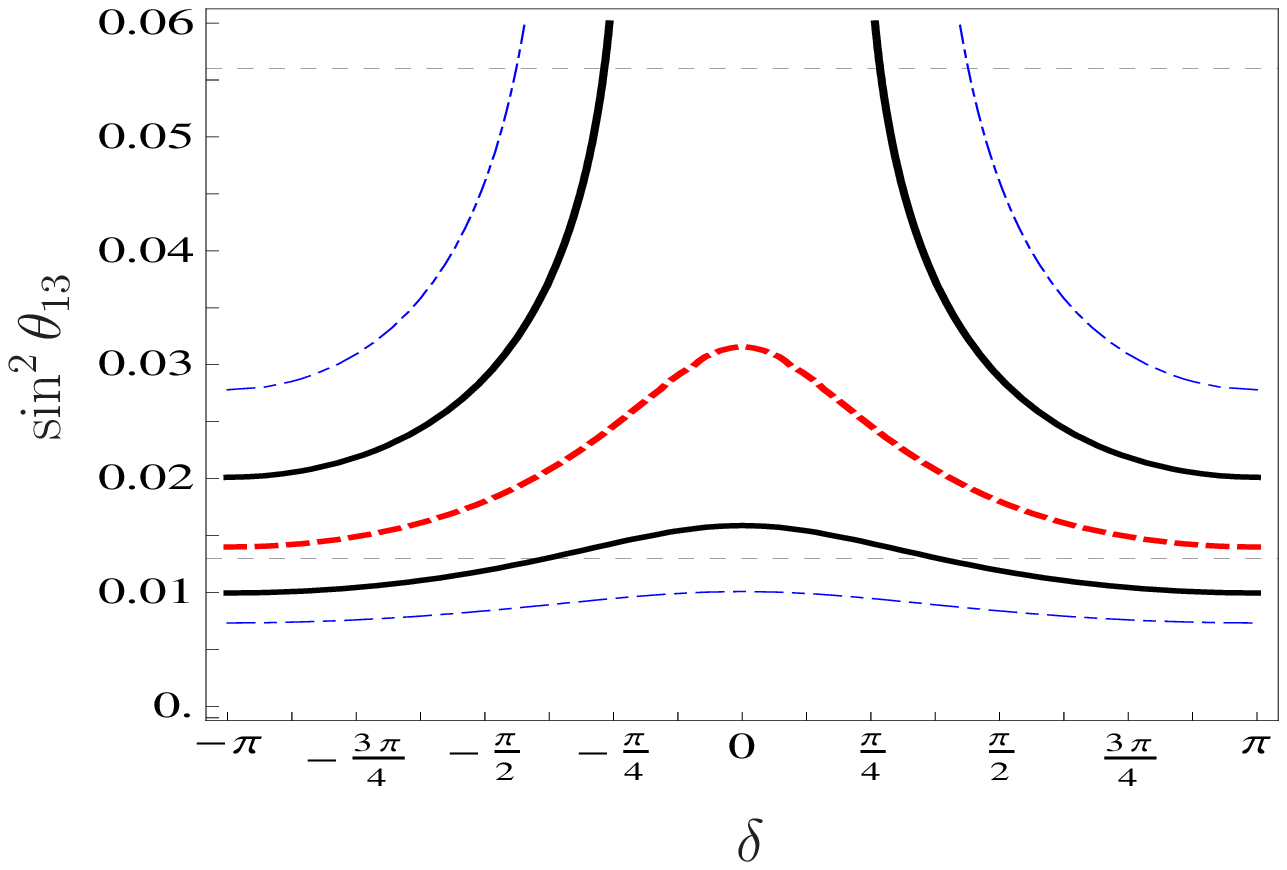}%\end{center}
\caption{\it Left panel: correlation among $\delta$ and $\sin^2 \theta_{13}$ as obtained in our model. 
The $1\sigma$ result, obtained  varying the other oscillation parameters also in their $1\sigma$ allowed ranges,
is showed with solid lines, whereas the $2\sigma$ result 
is showed with the dot-dashed line. The dashed line is the relation obtained when  
$\theta_{12}, \theta_{23}$ and $\alpha$ are fixed to their best fit values. Horizontal lines represent 
the upper limits on $\sin^2 \theta_{13}$ (upper dashed line) 
and the best fit values (lower dashed line) from \cite{data1}. Right panel: the same  as the left panel but assuming exactly diagonal charged lepton mass matrix.}
\label{t13-delta}
\end{figure}
\end{center}

In the left panel of Fig.\ref{t13-delta} we show the dependence of $\sin^2 \theta_{13}$ as a function of $\delta$ 
taking $\theta_{12}, \theta_{23}$ and $\alpha$ inside their experimental ranges. 
In particular, the solid line represents the $1\sigma$ correlation when also the other parameters are left free to 
vary in their $1\sigma$ allowed ranges quoted in \cite{data1}, whereas the $2\sigma$ correlation 
is represented by the dot-dashed line. Finally, dashed line is the relation obtained when  
$\theta_{12}, \theta_{23}$ and $\alpha$ are fixed to their best fit values. We also included 
the upper limit on $\sin^2 \theta_{13}$ at 3$\sigma$ (upper horizontal dashed line) and the best fit value
of ref.\cite{data1} (lower horizontal dashed line).
We can see that, even considering the $2\sigma$ uncertainty, the predicted values for  
$\sin^2 \theta_{13}$ are different from zero so that, to a very good accuracy, our model is compatible with
deviation from $\theta_{13}=0$ for any value of the CP violating phase. 
The precise value of $\theta_{13}$, however, relies on the assumed magnitude for $\delta$; in particular, 
the CP conserving case $\delta = 0$ is the most promising one to allow large $\theta_{13}$ 
(even above the current limits) whereas around $\delta\sim \pm \pi$ we get the smaller $\theta_{13}$ 
allowed in our model. It is interesting to observe that, in the case of maximal CP violation
\footnote{Maximal CP violation can be observed in incoming experiments T2K and NO$\nu$A, see for instance \cite{Nunokawa:2007qh,Huber:2009cw}.}
and for the 
other oscillation parameters to their best fit values, 
the predicted $\sin^2 \theta_{13}$ is fully compatible with the best fit value obtained in \cite{data1},
$\sin^2 \theta_{13}\sim 0.01$. Notice that, in the case of diagonal charged lepton mass matrix, the pattern of the $\theta_{13}-\delta$ correlation would have been quite similar, as it can be seen investigating the 
right panel of Fig.\ref{t13-delta}. The fact that the corrections coming from   $U_L$ in eq.(\ref{ul}) are as large as the values of $\sin^2 \theta_{13}$ is responsible for lowering the allowed  $\theta_{13}$ for $\delta\sim\pm \pi$.
For maximal CP violation at the best fit point, the Jarlskog invariant \cite{Jarlskog:1985cw} is as follows:
\bea
J=c_{12}s_{23}c_{13}^2s_{12}s_{23}s_{13}\sin \delta=0.023\,.
\eea

\begin{center}
\begin{figure}[h]\begin{center}
\includegraphics[width=10cm]{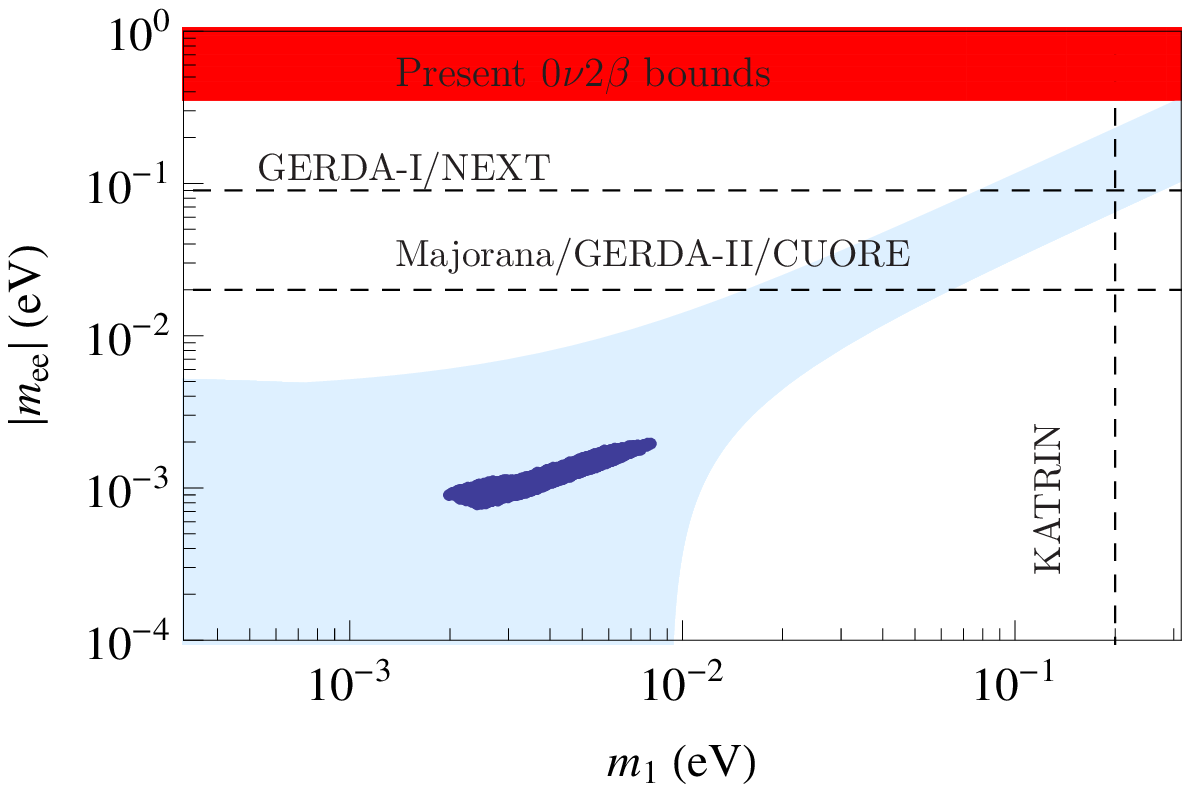}\end{center}
\caption{\it Model predictions for $|m_{ee}|$ as a function of the lightest neutrino mass $m_1$. 
We also show the allowed regions for the normal hierarchy (gray band). The two dashed horizontal lines represent the  
experimental sensitivity of some of the forthcoming experiments while the dashed vertical line is 
the upper limit for the sum of the absolute neutrino masses  from cosmological data. For references to experiments see~\cite{Osipowicz:2001sq,:2008qu,Smolnikov:2008fu,Giuliani:2008zz,Granena:2009it}.}
\label{mbb2}
\end{figure}
\end{center}

The next observable we want to discuss is the effective mass $m_{ee}$ entering in the neutrinoless double beta decay.
In the basis where the charged leptons are diagonal, $m_{ee}$ is nothing but the (11) element of the neutrino mass matrix. According to eq.(\ref{Mnu}), this should vanishes as long as the rotation in the charged leptons is proportional to the identity matrix. Since this is not the case, a non-vanishing  $m_{ee}$ is generated by the rotation (\ref{ul}) and it is expected to be  small because of the smallness of its off-diagonal entries.
This is what we can observe in Fig.(\ref{mbb2}),  where we plot the model predictions for $m_{ee}$ as a function of the lightest neutrino mass $m_1$.
For $m_1$ below ${\cal O}(10^{-2})$ eV we get $|m_{ee}| \sim 10^{-3}$ eV and then outside the range of future experimental sensitivities. We also see that the allowed range for the lightest neutrino mass 
is around $10^{-3}- 10^{-2}\, eV$; this is because, as already mentioned in the introduction, the Fritzsch texture gives a correlation 
between  $\theta^\nu_{12}$
\footnote{Note that  in the case of diagonal charged leptons the angle $\theta_{12}$ 
corresponds exactly to $\theta_{12}^\nu$ and therefore from eqs. (\ref{t12}) and (\ref{massrel1}), 
the solar mixing angle does not depend on the absolute scale of neutrino mass $m_1$, while in our 
case this relation acquires a small correction proportional to $\sqrt{m_e/m_\mu}$.}
and the ratio $m_1 /m_2 $ that, together with the two measured square mass differences,
fix the absolute neutrino scale in this range.

\section{Conclusion}
\label{conc}

In this paper we have studied a leptonic model based on the discrete $S_3$ 
permutation flavor symmetry. We extended the scalar sector of the Standard Model by 
introducing three more Higgs doublets and one scalar singlet. We have carefully studied the problem of the 
minimization of the potential and found that a solution of the form $(v,0)$ for the Higsses in the $S_3$  doublet representation is a viable 
minimum of the potential. With such a minimum, we obtain a two-zero Fritzsch-texture for the neutrino mass matrix 
and a  nearly diagonal and hierarchical charged lepton mass matrix. 
As a consequence of the two zeros of the  Fritzsch texture,
we get a strong correlation between the reactor angle $\theta_{13}$ and the Dirac CP phase $\delta$.
In particular, for $\delta\sim \pm\pi/2$ we predict  $\sin^2\theta_{13}\approx 0.01$, a value which is 
very close to the best fit value quoted in \cite{data1}. 
Beside the reactor angle, we also investigated the 
prediction for the effective mass $m_{ee}$ governing the rate of the 
$0\nu\beta\beta$ decay, founding
$m_{ee}\approx 10^{-3}\,eV$, one order of magnitude less than the
sensitivities of the future  experiments.

\section{Acknowledgments}
Work of SM and EP supported by the EC contract UNILHC PITN-GA-2009-237920,
by the Spanish grants FPA2008-00319 and CDS2009-00064 (MICINN)
and PROMETEO/2009/091 (Generalitat Valenciana)
and by European Commission Contracts
MRTN-CT-2004-503369 and ILIAS/N6 RII3-CT-2004-506222.
D.M. was supported by the Deutsche Forschungs-gemeinschaft, contract WI 2639/2-1. 
D.M. also acknowledges the AHEP Group of Valencia for their hospitality during the 
earlier stages of this work.

%\appendix
%\section{Deviation from tri-bimaximal}

\end{document}